\documentclass[prl,superscriptaddress,10pt,a4paper,sort&compress,twocolumn]{revtex4-2}
\usepackage{amsmath,amssymb,amsthm,mathtools,bm}
\usepackage{natbib}
\usepackage{graphicx}
\usepackage[inkscapelatex=false]{svg}
\usepackage{hyperref}
\usepackage[capitalise,nameinlink]{cleveref}
\usepackage{overpic}
\usepackage{siunitx}
\usepackage{soul}

\newcommand \bl{}

\newcommand{\cm}[1]{{\color{olive}#1}}
\newcommand{\pdiff}[3][]{\frac{\partial^{#1}{#2}}{\partial{#3}^{#1}}}
\newcommand{\intd}[1]{\,\mathrm{d}#1}
\renewcommand{\vec}[1]{\bm{#1}}
\newcommand{\changed}[1]{{#1}}
\newcommand{\changedd}[1]{{#1}}
\newcommand{\changeddd}[1]{{#1}}

\begin{document} 
\title{Minimal design of a synthetic cilium}


\author{Cl\'ement Moreau} 
\email{clement.moreau@cnrs.fr}
\affiliation{Nantes Universit\'e, \'Ecole Centrale Nantes, CNRS, LS2N, UMR 6004, F-44000 Nantes, France}
\affiliation{Research Institute for Mathematical Sciences, Kyoto University, Kyoto, 606-8502, Japan}

\author{Benjamin J. Walker} 
\email{bjw43@bath.ac.uk}
\affiliation{Department of Mathematical Sciences, University of Bath, Bath, BA2 7AY, UK}
\affiliation{Department of Mathematics, University College London, London, WC1E 6BT, UK}

\author{Rebecca N. Poon} 
\affiliation{Living Systems Institute \& Department of Mathematics and Statistics, University of Exeter, Exeter EX4 4QD, United Kingdom}

\author{Daniel Soto} 
\affiliation{ School of Physics, Georgia Institute of Technology, Atlanta, GA 30332, United States of America}

\author{Daniel I. Goldman} 
\affiliation{ School of Physics, Georgia Institute of Technology, Atlanta, GA 30332, United States of America}

\author{Eamonn A. Gaffney} 
\email{gaffney@maths.ox.ac.uk}
\affiliation{Mathematical Institute, University of Oxford, Oxford, OX2 6GG, United Kingdom}

\author{Kirsty Y. Wan} 
\email{k.y.wan2@exeter.ac.uk}
\affiliation{Living Systems Institute \& Department of Mathematics and Statistics, University of Exeter, Exeter EX4 4QD, United Kingdom}
\date{\today}

\begin{abstract}
We study a \changed{slender} filament beating in a viscous fluid with \changed{novel} curvature-dependent bending stiffness. Our numerical and experimental investigations reveal that such differential stiffness can sustain planar bending waves far along flexible filaments\changed{, in stark contrast to the uniform-stiffness case which requires more sophisticated control}. In particular, \changed{we establish basal actuation as} a viable, parsimonious mechanism for generating high-amplitude planar bending waves. \changed{Moreover}, the resulting beat patterns closely resemble the power-and-recovery strokes of propulsive biological filaments \changed{such as cilia}, suggesting \changed{extensive} applications in robotic and engineered systems.

\end{abstract}

\maketitle


\section{Introduction}
\changed{The cilium is a complex, highly conserved organelle found in diverse organisms where it contributes to a host of important functions, including flow manipulation and self-propulsion of microorganisms \cite{mitchell2017evolution,brennen1977}.
Inspired by these natural capabilities, there has been significant interest and progress in the field in fabricating micro or even nanoscopic devices that can also be used to enable fluid pumping, swimming, particle mixing and manipulation.
Synthetic systems are currently unable to match the versatility of biological cilia, which has driven the development of sophisticated artificial control mechanisms. Magnetic actuation has become the favoured strategy to precisely engineer realistic beat patterns in ensembles of artificial cilia, though many other methods have been employed, including pneumatic, electrochemical, and optical actuation  \cite{ulislam2022microscopic,gu_magnetic_2020}. 
Rather than devise increasingly sophisticated modalities of control, we consider an alternative approach: is it possible to engineer the material properties of a filament to produce realistic, cilia-like beat patterns using only a simple control mechanism?

}




\begin{figure}
    \centering
    \includegraphics[width=0.4\textwidth]{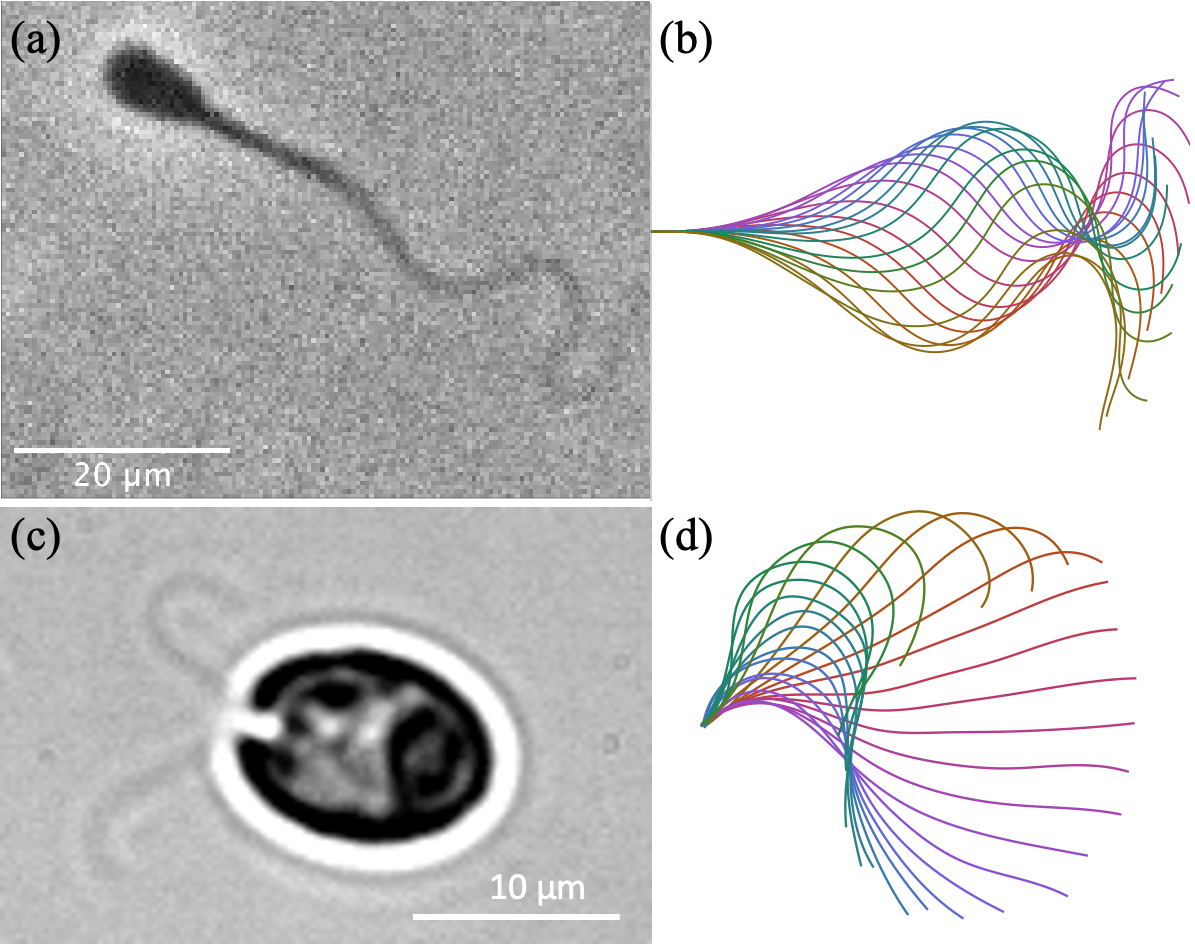}
    \caption{Examples of flagellar actuation in biological micro-swimmers. Beating flagella of (a) a bovine spermatozoon and (c) \textit{Chlamydomonas reinhardtii} display large amplitude bending at their distal ends. (b,d) Traces of the waveforms, color-coded in time, highlight the symmetric beat pattern for the sperm cell and marked asymmetry in \textit{Chlamydomonas} between the power and recovery strokes. Images (a,b), (c,d) are reproduced with permission from \citet{walker2020} and \citet{wan2014}, respectively.}
    \label{fig: bio examples}
\end{figure}
 
\changed{To explore this idea, we revisit} a seminal study of the fluid-structure interactions of a swimming mammalian spermatozoan, \cref{fig: bio examples}a from the 1950s. Here, \citeauthor{machin1958} contrasted two popular competing hypotheses of the time \citep{machin1958}: (1) \changed{the filament is driven from the base} (at its interface with the cell body) but is otherwise passive, and (2) \changed{it} is driven by active elements along its length. Through a theoretical analysis of an idealised Euler elastica in the absence of inertia, \citeauthor{machin1958} concluded that passive, proximally driven filaments were incapable of reproducing large-amplitude bending waves that are characteristic of eukaryotic cilia and flagella \cite{brennen1977,woolley2001,gallagher2019rapid}. In particular, \citeauthor{machin1958} noted that, depending on parameters, one of two things occurred: (1) the amplitude of bending waves decayed rapidly along the flagellum, so that distal amplitudes were effectively negligible (Fig 3 F-J of \cite{machin1958}), or (2) bending only occurred with long wavelength, with fewer than approximately 1.5 wavelengths being present along the flagellum (Fig 3 A-E of \cite{machin1958}).

\citeauthor{machin1958}'s study refuted the first of the two hypotheses for flagellar beating, a result later confirmed by experiments that culminated in the identification of \changed{spatially distributed dynein} as the molecular motors responsible for force generation inside these filaments \cite{gibbons1963,gibbons1965,summers1971}. 
\changed{As these filaments are remarkably conserved across eukaryotes (\cite{mitchell2017evolution,lindemann2022,merchant,gibbons1965} and \cref{fig: bio examples}), it is now well-established that ciliary beating is universally driven by distributed activity rather than localised actuation.} Indeed, cilia completely isolated from the cell body continue to beat in the presence of ATP \cite{geyer2022ciliary,bessen1980calcium,wan2022beat}. Hence, proximal actuation has been duly passed over as a driver of motility in eukaryotic systems. This notably contrasts with flagellated bacteria and archaea, which self-propel using a rotary motor that generates basal torque to spin a helical flagellum  \cite{berg1973bacteria,beeby2020propulsive}.

\changed{Thus, the prospect of driving cilia-like beating patterns in elastic filaments via basal actuation alone would appear to be unrealistic. However, while} the material properties of biological cilia and flagella are restricted by developmental or evolutionary constraints, engineered synthetic filaments may be designed with more freedom \cite{diaz2021minimal,ulislam2022microscopic,sareh2013swimming,Lim2023}. 
This raises a simple question: by modifying the mechanical properties of a filament, can the conclusions of \citeauthor{machin1958}'s analysis be sidestepped? Can we produce realistic bending waves in engineered passive filaments for low-Reynolds number propulsion using only basal actuation? If so, this simple mechanism would facilitate the development of markedly simple synthetic systems \changed{without intricate mechanisms of distributed driving or control. In turn, these may be designed to  exploit asymmetric ciliary beating either for propulsion, such as in multi-ciliated swimmers like \textit{Chlamydomonas}, or for flow generation, as observed in epithelial cilia.}  \changed{This strategy is analogous to implementing directional compliance} in \changedd{robotic appendages} modelled upon the limbs of insects \cite{spagna2007distributed}, the undulatory motion of snakes \cite{wang2020directional} \changedd{or magnetic metamaterials \cite{wu2019symmetry,obayashi2023control,mohaghar2024effects}}, which can \changed{greatly reduce the costs of gait computation.} 

\changed{Seeking to embody novel functionality within simple synthetic systems, we combine theoretical study with a novel engineered physical system. In this Letter we establish how sustained, cilia-like beating patterns can be driven by basal actuation alone in appropriately designed filaments.}

\section{Machin's low-amplitude model}

\changed{To motivate our eventual design principle, we consider} (following \cite{machin1958}) the propagation of bending waves along an elastic filament in a fluid using the dimensionless small-amplitude beam equation
\begin{equation}\label{eq: small amplitude filament}
    \pdiff{y}{t} = -\pdiff[2]{}{x}\left(E\pdiff[2]{y}{x}\right)
\end{equation}
with boundary conditions
\begin{subequations}
\begin{gather}
    y(0,t) = 0\,, \quad \pdiff{y}{x}(0,t) = A\sin{t}\,,\\
    \pdiff[2]{y}{x}(1,t) = 0\,, \quad \changed{\pdiff[3]{y}{x}(1,t)=0}\,.
\end{gather}
\end{subequations}
This system corresponds to a filament with displacement $y$ driven by rotation about a hinge at $x=0$, with moment-free and shear-free conditions at the tip $x=1$. Throughout, we use dimensionless quantities such that $E$ captures the ratio of bending resistance to drag, the driving oscillations occur at unit frequency, and the filament is of unit length. Balancing terms in \cref{eq: small amplitude filament} gives rise to a natural dimensionless lengthscale $\changed{l\propto\sqrt[4]{E}}$, over which planar elastic waves decay exponentially.

Many studies have generalised \citeauthor{machin1958}'s model, going beyond the small-amplitude limit and into three dimensions \cite{brokaw1966,hines1978,camalet2000,Rallabandi2022},  
or have considered \changed{hydrodynamic and basal synchronisation of several flagella \cite{guo2018bistability,guo2021intracellular}, follower force actuation \cite{man2019morphological}} or non-uniform, spatially-varying stiffness \cite{peng2017maximizing,neal2020doing}.  Here, \changed{we seek an alternative generalisation and} incorporate state dependence into the stiffness, setting $E=E(\pdiff[2]{y}{x})$ to depend on the local linearised curvature. For simplicity, we consider $E$ to be piecewise constant, so that
\begin{equation}
    E(\changeddd{\kappa}) = \begin{cases}
        E_L & \text{if } \changeddd{\kappa} > 0\,,\\
        E_R & \text{if } \changeddd{\kappa} \leq 0
    \end{cases}
\end{equation}
for unequal constants $E_L > E_R > 0$ \changeddd{and curvature $\kappa\in\mathbb{R}$}. Solving \cref{eq: small amplitude filament} is now somewhat involved, but we can estimate the appropriate decay lengthscale by considering a further simplification. 
 
Suppose that, instantaneously, the filament is divided into two equal regions of positive and negative curvature, similar to a sinusoid over a single period. \changed{Up to constants and over a fixed timescale, the appropriate decay lengthscale} in the region of positive curvature is $l_L = \sqrt[4]{E_L}$; in the region of negative curvature, the decay lengthscale is $l_R = \sqrt[4]{E_R}$. Note $l_L > l_R$. Hence, the amplitude of planar bending waves after passing through both regions experiences decay scaling with $\exp{(-\frac{1}{2}[1/l_L + 1/l_R])}$, as opposed to either of $\exp{(-1/l_L)}$ and $\exp{(-1/l_R)}$. Hence, there is an effective decay lengthscale of $2(1/l_L + 1/l_R)^{-1}$ in the filament, equal to the harmonic mean of the two individual lengthscales of decay. Notably, this is at least as large as the smallest lengthscale $l_R$, and approaches $2l_R$ as $E_L\to\infty$. Intuitively, this latter case corresponds to a filament that is extremely hard to bend in one direction, whilst being relatively compliant in the other. 

This reasoning readily generalises to filaments where a proportion $\alpha\in(0,1)$ is bent such that $E=E_R$. In this case, the effective lengthscale is $([1-\alpha]/l_L + \alpha/l_R)^{-1}$, which approaches $l_R/\alpha > l_R$ as $E_L\to\infty$. Whilst clearly $\alpha$ should not be considered to be a fixed quantity in a dynamic filament, this back-of-the-envelope scaling analysis suggests that state-dependent bending stiffness could greatly enhance the effective decay length, with filaments thereby being able to transmit bending waves far along their lengths. Importantly, this mechanism unifies desirable properties of both high and low stiffness regimes: bending waves can be of high curvature (associated with the lower bending stiffness) and also propagate far along the filament (associated with the higher bending stiffness). Thus, filaments with differential stiffness are qualitatively different to their constant-stiffness counterparts.

\begin{figure}
    \centering
    \begin{overpic}[permil,width=0.47\textwidth]{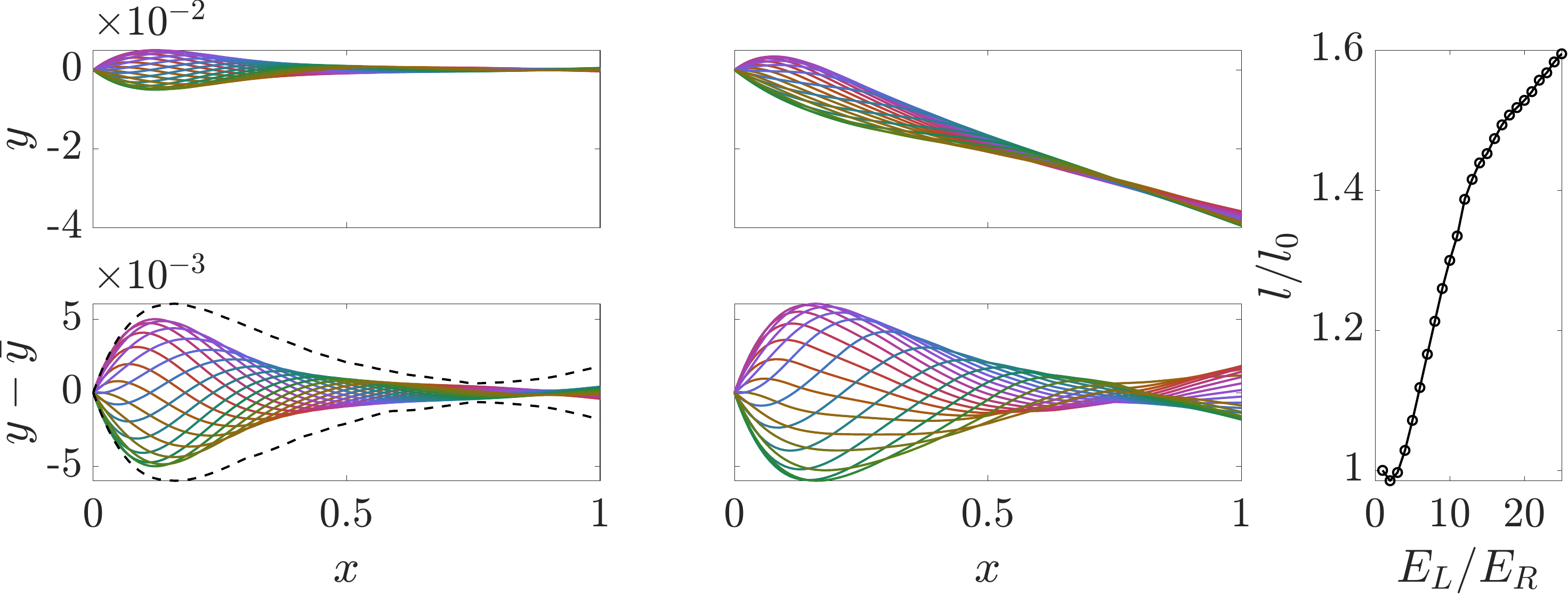}
    \scriptsize
    \put(-20,361){(a)}
    \put(410,361){(b)}
    \put(-20,190){(c)}
    \put(410,190){(d)}
    \put(792,361){(e)}
    \end{overpic}
    \caption{Differential stiffness and small-amplitude beating. (a,b) Computed solutions of the small-amplitude beam equation over time with (a) $E_L=E_R=10^{-4}$ and (b) $E_L=10^{-3}$, $E_R=10^{-4}$. (c,d) Deviation from the average configuration $\bar{y}$, highlighting increased amplitude with differential filament stiffness. The envelope of (d) is overlaid on (c) as black dashed curves. (e) Estimated decay lengthscale $l$ as a function of differential stiffness, relative to the lengthscale $l_0$ in the equal-stiffness case. Here, $A=0.1$.}
    \label{fig: small amplitude filament}
\end{figure}

To examine these idealised arguments thoroughly, a numerical study of filaments with differential bending stiffness is presented in \cref{fig: small amplitude filament}, computed using FEniCS \cite{Fenics}, wherein the state-dependent stiffness is taken to vary smoothly (but rapidly) between $E_L$ and $E_R$ (detailed in the Supplementary Material). We encourage readers to explore this system for themselves via a \href{https://visualpde.com/nonlinear-physics/nonlinear-beams.html}{browser-based VisualPDE simulation} \cite{walker2023visualpde}. The contrast between a uniform filament (a) and a filament with state-dependent stiffness (b) is marked, with the latter being highly asymmetric. The amplitude of distal motion is significantly greater in (b) than in (a), visible in the mean-centered plots of (c) and (d). This enhancement of both the distal amplitude and the apparent decay lengthscale agree qualitatively with the scaling analysis. This is further supported by numerical estimation of the decay lengthscale as a function of the degree of differential stiffness (\cref{fig: small amplitude filament}e), described in the Supplementary Material. Thus, at least in the small-amplitude regime, state-dependent stiffness is a plausible mechanism for propagating planar bending waves along filaments.


\begin{figure}
    \centering
    \begin{overpic}[permil,width=8.5cm]{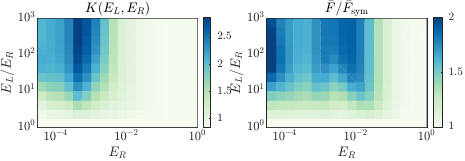}
    \put(0,330){(a)}
    \put(500,330){(b)}
    \end{overpic}
    \begin{overpic}[permil,width=8.6cm]{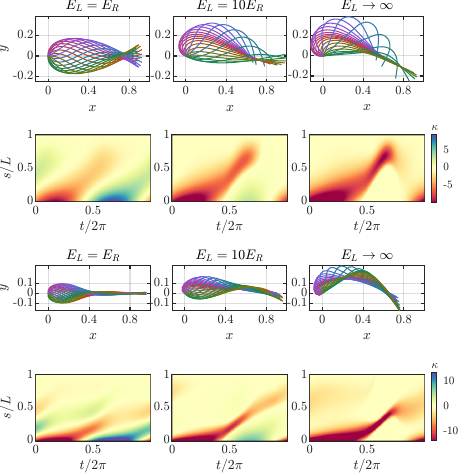}
    \put(0,980){(c)}
    \put(0,460){(d)}
    \end{overpic}
    \caption{Effects of differential stiffness in the geometrically nonlinear regime ($A=\pi/2$), obtained from numerical solution of \cref{eq: large amplitude filament}. (a,b) Relative increase in (a) maximal curvature and (b) average force at the pinned end, for a range of stiffnesses $E_R$ and differential stiffnesses $E_L/E_R$. 
    Example waveforms are shown below for (c) medium stiffness ($E_R = 1.6 \times 10^{-3}$) and (d) low stiffness ($E_R = 1\times 10^{-4}$). (c,d) Example beating patterns (rotated in the beating plane to align horizontally) and associated curvature kymographs for filaments of (c) medium stiffness ($E_R=\SI{1.6e-3}{}$) and (d) low stiffness ($E_R=\SI{1e-4}{}$). From left to right: symmetric beating ($E_L=E_R)$, resembling the classical solution of \cref{fig: small amplitude filament}; an intermediate case $(E_L/E_R=10)$, matching that studied in the small-amplitude regime; and a highly asymmetric ``locking'' case $(E_L/E_R=10^3)$, effectively forbidding negative curvatures.
    }
    \label{fig: nonlinear filament}
\end{figure}

\section{Large-amplitude bending}
To investigate beyond the small-amplitude regime, we adopt a geometrically nonlinear framework \cite{moreau2018,walker2019}. The analogue of \cref{eq: small amplitude filament} in the geometrically nonlinear regime is
\begin{equation}\label{eq: large amplitude filament}
    E\pdiff{\theta}{s} + \int\limits_s^1 [\vec{x}(\tilde{s}) - \vec{x}(s)]\times \vec{f}(\tilde{s}) \intd{\tilde{s}} = 0
\end{equation}
written in terms of the tangent angle $\theta$ (measured relative to a fixed axis), position $\vec{x} = (x,y)$, and hydrodynamic drag $\vec{f}$, each functions of arclength $s\in[0,1]$ and dimensionless. The derivation of this integral equation and the underlying assumptions are described in detail by \citet{moreau2018}, recounted in part in the Supplementary Material. In the nonlinear regime, curvature is $\kappa = \pdiff{\theta}{s}$, so that $E=E(\kappa)$. Solving this integro-differential equation numerically (as described in the Supplementary Material) allows us to test the predictions of the linear theory well beyond its regime of validity. In the following, we set $A=\pi/2$.

In this general regime, a quantitative measure of the effects of stiffness is the relative increase in maximum curvature over one beating period, averaged over the flagellum, defined for given $E_L$ and $E_R$ as
\begin{equation}
    K(E_L,E_R) = \int_0^1 \frac {\max_{t} | \kappa_{E_L,E_R}(s,t) |}{\max_{t} {| \kappa_{E_R,E_R}(s,t)} | }\intd{s},
    \label{eq:relative increase}
\end{equation}
where the maximum over $t$ is computed over one time period in the limiting periodic regime. Naturally, $\kappa_{E_L,E_R}(s,t)$ and $\kappa_{E_R,E_R}(s,t)$ refer to the curvature for the asymmetric and symmetric filaments, respectively.

This quantity is presented in panel (a) of \cref{fig: nonlinear filament}, plotted as a function of $E_R$ and the differential stiffness $E_L/E_R$, demonstrating that differential stiffness can greatly enhance the curvatures attained by a filament during a beating period. The relative increase is three-fold around $E_R \approx 1 \times 10^{-3}$.

To confirm that the increase in average curvature is localised towards the distal end of the filament, we also determined the maximum value $K_{\max}$ and its argument $s_{\max}$ over $s \in [0,1]$ of the integrand in \cref{eq:relative increase}, indicating where the qualitative change between uniform and differential stiffness is the greatest. For any value of $E_L$ and $E_R$, $s_{\max}$ takes values between 0.5 and 0.8, confirming the trend in \cref{fig: nonlinear filament}. Around $E_R \approx \SI{1e-3}{}$, the corresponding value of $K_{\max}$ goes up to $6$, further highlighting the capacity of differential stiffness to transport and even amplify curvature for large beating amplitudes. 

\citeauthor{machin1958} identified (in the small-amplitude limit) that exponential decay significantly hinders planar wave propagation towards the distal end when the filament length exceeds five times the characteristic lengthscale $\sqrt[4]{E_R}$. Surprisingly, the parameter regime in which we see maximal curvature increase due to differential stiffness approximately corresponds to this case. Moreover, the curvature increase remains markedly high even when considering very small $E_R$ ($<10^{-4}$, representing a filament more than 10 times longer than the relaxation lengthscale), for which the effect of proximal actuation with uniform stiffness is virtually negligible at the distal end. 

\Cref{fig: nonlinear filament}b displays the magnitude $\bar{F}$ of the pinning force, exerted on the point of attachment at the proximal end, averaged over one beating period, and normalised by the time-averaged force without differential stiffness, $\bar{F}_{\mathrm{sym}}$. As should be expected, maintaining a prescribed actuation while increasing filament stiffness for positive curvature comes at the cost of increasing the average pinning force. However, this relative increase remains smaller than the relative curvature enhancement through most values of $E_R$ and $E_L$, indicating that differential stiffness constitutes an efficient way of increasing curvature according to this measure. For all values of $E_L/E_R$, the relative force $\bar{F}/\bar{F}_{\mathrm{sym}}$ has a local minimum with respect to $E_R$ for $E_R \approx \SI{1.6e-3}{}$. Simultaneously, $K(E_L,E_R)$ is approximately maximal, so that the ratio between $K(E_L,E_R)$ and $\bar{F}/\bar{F}_{\mathrm{sym}}$ is also approximately maximal. This value of $E_R$ appears to be optimal for enhanced curvature generation with differential stiffness. 

Panels (c) and (d) of \cref{fig: nonlinear filament} illustrate filament beats in the large amplitude regime for various degrees of differential stiffness. In the left column, we take $E_L = E_R$ and reproduce the symmetric, short-lengthscale beating predicted by \citeauthor{machin1958}'s low-amplitude analysis for differing baseline stiffness $E_R$. 
In the subsequent columns, increasing degrees of differential stiffness yield larger amplitude deviations at the distal end of the filament while also allowing for high-curvature planar bending waves to propagate. 
These beats are further represented in the plots of signed curvature $\kappa$ beneath the beating patterns, from which the high curvatures of the asymmetric regimes are evident. The spatial extent of bending waves is also clearly visible, with differential stiffness giving rise to far-reaching bending. Moreover, in the case $E_R \approx \SI{1.6e-3}{}$ (\cref{fig: nonlinear filament}c), the resulting beat patterns begin to resemble the \emph{power-and-recovery strokes} typical of most cilia, including \textit{Chlamydomonas} (\cref{fig: bio examples}d), other protists \cite{Wang2020,sleigh1989ciliary,laeverenz2024bioelectric}, and vertebrate respiratory cilia \cite{Sanderson1981}.

Thus, even beyond the small-amplitude regime, differential stiffness presents itself as an \changed{effective} mechanism for enhancing elastic planar wave propagation. 

\begin{figure}
    \centering
    \includegraphics[width=8.6cm]{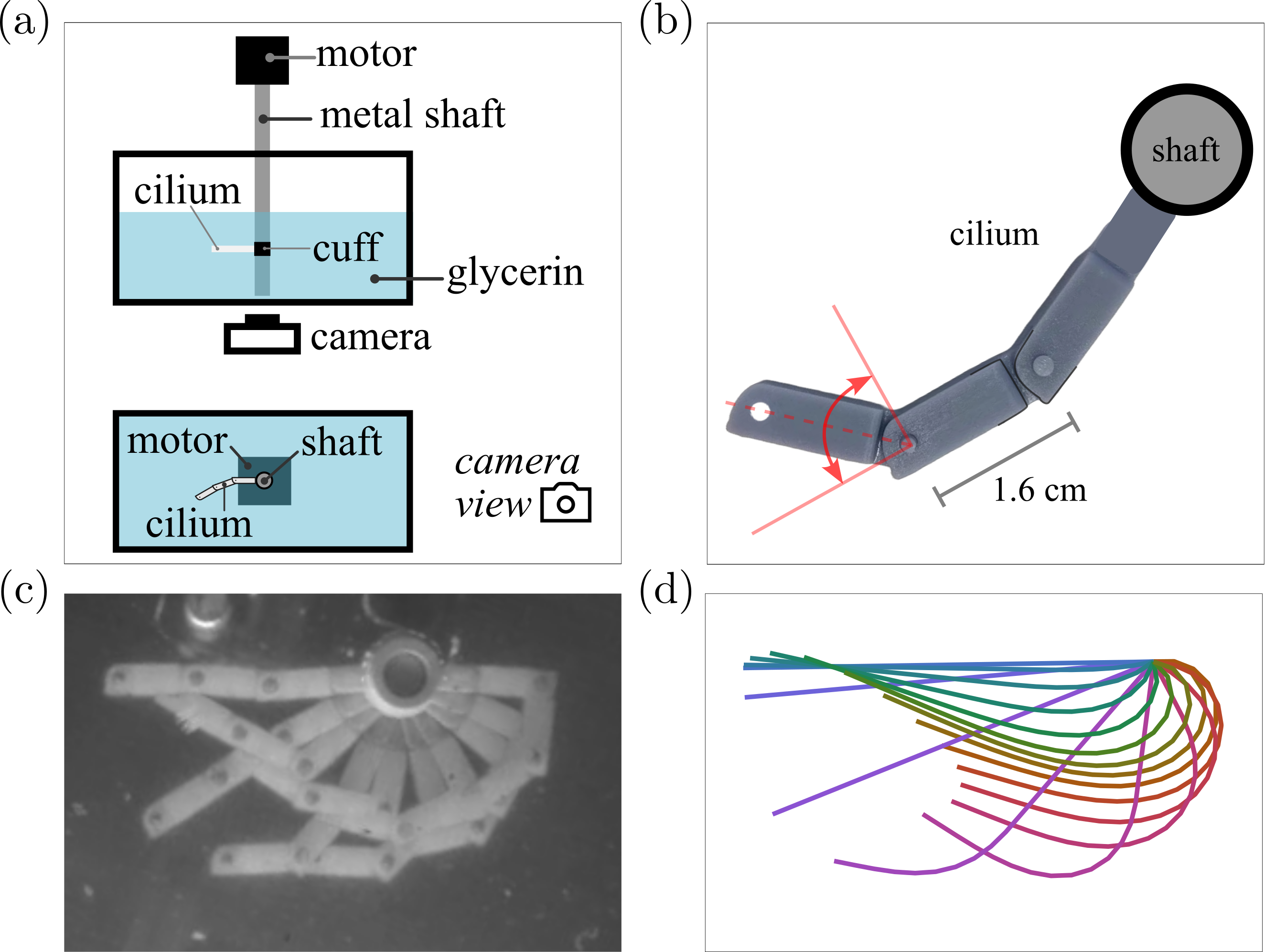}
    \caption{Experimental realisation of differential stiffness in a bio-inspired macroscale robotic cilium \changed{(Supplementary Video 1)}. Motion is driven only by actuating the base of the filament. (a) Diagram of the experimental setup, not to scale. The cilium is placed in a tank of glycerin, and imaged from below. The cilium is connected to the driving motor by a vertical metal shaft.  
    (b) Detail of the locking mechanism. The links are designed so that each link can rotate freely only within the \SI{90}{\degree} angle indicated in red.  (c) Snapshots of the robotic filament overlaid in time compared to (d) a simulated waveform, with $A=\pi/2,$ $E_R = \SI{2e-3}{}$, $E_L\gg E_R$ highlighting marked agreement between the synthetic and \textit{in silico} systems.
    }
    \label{fig: experiment realisation}
\end{figure}
\section{Engineering cilia-like beating}
Given the striking resemblance of the simulated beat of \cref{fig: nonlinear filament}c with those seen in certain biological specimens, \changed{we now ask if we can} realise biomimetic beating in practice using only basal actuation and differential stiffness?
To explore this, we built a macroscopic robotic realisation of a basally actuated filament, schematised in \cref{fig: experiment realisation}a. The continuum limit is difficult to implement \changed{in practice}, so our robotic filament comprises a finite number of links connected by custom hinges that are designed to qualitatively mimic differential stiffness \cite{diaz2021minimal}. 

These hinges, illustrated in \cref{fig: experiment realisation}b, are asymmetric in shape, so they only admit bending in one direction, similar to the ``locking'' limit $E_L \gg E_R$ explored above. Each of the three links is \SI{1.6}{\cm} long, with a \SI{1.2}{\cm} basal attachment, so that the total length of the filament from its rotation point is \SI{6}{\cm}. The locking hinges are designed to reproduce the most extreme case of differential stiffness in the limit of $E_L \gg E_R$ (as seen in \cref{fig: nonlinear filament}c). \changed{Notably, these hinges are simple to customise and produce.}

The filament is driven at the base by a XL430-W250 dynamixel motor, which rotates through \SI{180}{\degree} at an angular velocity of \SI{5}{\degree\per\second}. 
To remain in the low-Reynolds regime, the filament is placed in a tank of glycerin, which has density of \SI{1.26e3}{\kg\per\m\cubed} and viscosity \SI{1.4}{\Pa\s}. The oscillatory Reynolds number $\rho f L^2/\eta$ = \changedd{0.3}. Visual inspection of the beating filament confirmed that inertial coasting was negligible. The tank measures \SI{60}{\cm} by \SI{32}{\cm}, with depth \SI{18}{\cm}. The filament is positioned far from walls to minimise boundary interactions. 
Imaging was performed at \SI{1}{\Hz} using a Logitech webcam. The cilia were resin printed using an Anycubic Photon Mono X, in Anycubic resin, and density matched to the glyerin.

Prescribing a basal amplitude $A=\pi/2$, the beat pattern of the robotic filament is showcased in \cref{fig: experiment realisation}c. 
Even without elasticity, the power and recovery strokes are clearly distinguished, with experimental waveforms closely resembling the cilia-like beating observed in our nonlinear simulations (\cref{fig: experiment realisation}d). \changedd{Further analysis, including a maximum intensity projection of the full video (Supplementary Video 1, 12 beat cycles in total), shows that small tracer particles in the fluid move a net distance of approximately \SI{0.1}{\cm} per beat cycle, evidencing net pumping of fluid by this artificial cilium. (See the Supplementary Material for details.)} A full and faithful model of the robotic system and a complete parameter analysis will be explored in future study. 

\section{Summary and conclusions}



 
\changed{Through a combination of experiments and theory, we have demonstrated that enhanced wave propagation can be achieved with marked simplicity, realising cilia-like beats in engineered filaments using basal actuation alone, \changedd{contrasting for instance with distributed magnetic actuation in recent similar designs of robots with structural asymmetric stiffness \cite{wu2019symmetry,mohaghar2024effects}}. 
By incorporating a curvature-dependent stiffness we pragmatically generate biologically realistic planar waveforms using only a simple modality of control. This highlights and newly establishes the ready potential for simple, parsimonious synthetic systems to reproduce this well-studied, broadly applicable and widely conserved biological phenomenon.
More generally, it paves the way for new avenues of multidisciplinary enquiry into the motility and actuation of synthetic systems at the microscale.

{\bl It is well known that mammalian sperm flagella have a spatially heterogeneous bending stiffness with higher proximal stiffness \cite{leisch2004,leisch2008}. This is potentially analogous to the larger initial link  in the macroscale robotic cilium  depicted in  Fig.~\ref{fig: experiment realisation}b, which prevents proximal bending, and the functional  impact of time-independent axisymmetric heterogeneity has  been explored in a number of studies \cite{lindemann1996,gadelha2019}. However,} biological cilia also possess  numerous protein complexes positioned along axonemal microtubule arrays that could harbour {\it dynamic} structural heterogeneities {\bl that do not possess axisymmetry,}  contributing to an effective state-dependent stiffness \cite{bui2012polarity,dutcher2020asymmetries,chen2023situ,Nicastro2018}. The potential for dynamic tuning of \changedd{apparent {\bl and state dependent}} bending stiffness in these structures has not been previously considered. Nonetheless, the dynamic modulation of these asymmetries in live cells, e.g. by phosphorylation, could contribute to the generation of distinct beat patterns on the same structure, noting that flagella stiffness changes with the level of ATP \cite{Lindemann1973,Okuno1979,Okuno1980}, which in turn has  been associated with the extent of phosphorylation of outer dynein arm-linked proteins \cite{Yoshimura2007}. In particular, to achieve control systems} for microscale movement, cells must exploit the mechanical or physical intelligence of their bodies, which could be \changedd{realised, at least in part,} through state-dependent bending stiffness of their cilia.

\section*{Acknowledgements}
This work was funded by the Japan Society for the Promotion of Science (Fellowship No. PE22023 and Grant No. 22KF0197 to C.M) and the Research Institute for Mathematical Sciences, an International Joint Usage/Research Center located at Kyoto University (C.M), by the Royal Commission for the Exhibition of 1851 (B.J.W), by the European Research Council under the European Union's Horizon 2020 research and innovation programme grant 853560 EvoMotion (K.Y.W), and a Company of Biologists travelling fellowship (R.N.P and K.Y.W).

For the purpose of Open Access, the author has applied a CC BY public copyright licence to any Author Accepted Manuscript (AAM) version arising from this submission.
 
\bibliography{bib}

\end{document}


\title{Minimal design of a synthetic cilium\\Supplementary Material}
\author{Cl\'ement Moreau, Benjamin J. Walker, Rebecca N. Poon, Daniel Soto,\\ Daniel I. Goldman, Eamonn A. Gaffney, Kirsty Y. Wan}
\date{}

\maketitle

\section{Numerical methods}

\subsection{Low-amplitude model}
We seek to numerically solve the low-amplitude beam equation
\begin{equation}\label{eq: small amplitude filament}
    \pdiff{y}{t} = -\pdiff[2]{}{x}\left(E\pdiff[2]{y}{x}\right)
\end{equation}
with boundary conditions
\begin{subequations}
\begin{gather}
    y(0) = 0\,, \quad \pdiff{y}{x}(0) = A\sin{t}\,,\\
    \pdiff[2]{y}{x}(1) = 0\,, \quad \pdiff[3]{y}{x}(1)=0\,,
\end{gather}
\end{subequations}
where $E=E(\pdiff[2]{y}{x})$ depends on the linearised curvature. We take the functional form of $E$ to be a smooth function that rapidly transitions between unequal constants $E_L > E_R > 0$, specifically
\begin{equation}\label{eq: stiffness}
    E(\sigma) = E_R + \frac{[E_L - E_R][1 + \tanh(\sigma/\epsilon)]}{2}
\end{equation}
for $\epsilon = 0.1$, which captures rapid transition over relevant curvatures. This PDE is solved in weak form using FEniCS \cite{Fenics} with adaptive implicit Euler timestepping and Lagrange elements. Source code is freely available at \url{https://github.com/Mar5bar/state-dependent-stiffness}.

An estimate for the decay lengthscale is computed from the numerical solutions as follows: for each discretised $x\in[0.2,0.6]
$, we computed the maximum absolute displacement from the temporal mean, with the limited range reducing the impact of end effects. These measurements of amplitude were fitted to an exponential of the form $ce^{-x/l}$ using a logarithmic transformation and a linear-least-squares residual, yielding an estimate for the decay lengthscale $l$.

\subsection{VisualPDE simulation}
The same equations as above are solved using VisualPDE.com via the numerical scheme described in \cite{walker2023visualpde}, though with parameters adjusted to aid visualisation. Note that this simulation is intended only to provide a qualitative analogue of the high accuracy simulations of the previous section.

\subsection{Large-amplitude model}
The development of the large-amplitude modelling framework is very similar to that of 
\citet{hall-mcnair2019a,walker2019,walker2021} and especially \citet{moreau2018}, which approximate a continuous elastic filament by a collection of links in a coarse graining of continuum elasticity. We modify the formulation of \citet{moreau2018} to include an actuated base and a state-dependent stiffness following \cref{eq: stiffness}, but otherwise precisely adopt their methodology. We direct the interested reader to the noted publications for a full account of the approach, including the constitutive assumptions, and summarise it briefly below.

\citeauthor{moreau2018}'s approach discretises the beam into $N$ straight segments of equal length and tracks the evolution of the orientation of each of these segments over time, relaxing the low-amplitude constraint inherent in the above formulation. The governing equations remain those of moment balance, with the curvature captured as $\pdiff{\theta}{s}$ in place of $\pdiff[2]{y}{x}$ in the geometrically nonlinear regime. Explicitly, the dimensional analogue of \cref{eq: small amplitude filament} is
\begin{equation}\label{eq: large amplitude filament}
    E\pdiff{\theta}{s} + \int\limits_s^1 [\vec{x}(\tilde{s}) - \vec{x}(s)]\times \vec{f}(\tilde{s}) \intd{\tilde{s}} = 0\,,
\end{equation}
where $\vec{f}(s)$ is the hydrodynamic drag associated with the motion at arclength $s$ and $E = E(\pdiff{\theta}{s})$. The drag is related to the motion of the filament via the resistive force theory of \citet{gray1955,hancock1953}, which reads
\begin{equation}
    \vec{f} = -C_N(\dot{\vec{x}}\cdot\vec{e}_N)\vec{e}_N -C_T(\dot{\vec{x}}\cdot\vec{e}_T)\vec{e}_T\,.
\end{equation}
Here, $\vec{e}_T(s)$ and $\vec{e}_N(s)$ are unit tangent and normal vectors to the filament and $\dot{\vec{x}}(s)$ is the velocity of a material point on the filament as measured in the laboratory frame. The positive constants $C_T$ and $C_N$ capture the magnitudes of hydrodynamic drag in the tangential and normal directions, respectively, and satisfy $C_N/C_T=2$.

To proceed numerically, \cref{eq: large amplitude filament} is applied at discrete arclengths $s_i = iL/N$ for $i\in\{0,\ldots,N-1\}$. Noting that the position $\vec{x}(s_i)$ can be written solely in terms of the base $\vec{x}(0)$ ($\vec{0}$ without loss of generality) and the segment orientations $\theta_i=\theta(s_i)$, this generates a closed system of $N$ ordinary differential equations in the unknown $\theta_i$. These equations are solved numerically in MATLAB \cite{Shampine1997} to yield the filament motion, utilising adaptive implicit timestepping with error tolerances set below $10^{-6}$. More concretely, the equations can be summarised in dimensionless form as 
\begin{equation}
    \mathrm{Sp}^4\bm{M}\dot{\vec{\theta}} = \vec{R}\,,
\end{equation}
where $\bm{M}$ is an invertible linear operator, $\dot{\vec{\theta}}$ is the time derivative of the vector of the unknown $\theta_i$, and $\vec{R}$ encodes the contribution of the state-dependent bending moments. The explicit and somewhat intricate form of these terms is given by \citet{moreau2018,walker2019}. Here, having implicitly non-dimensionalised, $\mathrm{Sp}$ is the governing dimensionless elastohydrodynamic parameter (the so-called \emph{sperm number}), defined via
\begin{equation}
    \mathrm{Sp}^4 = \frac{C_N L^4}{E_mI T}
\end{equation}
for typical bending modulus scale $E_m$, moment of inertia $I$, filament length $L$, and oscillation timescale $T$. In general, small $\mathrm{Sp}$ correspond to stiff filaments, whilst large $\mathrm{Sp}$ bend more readily in the fluid medium. In our study, we use a dimensionless parameter $E_R$ which is related to the sperm number via $\mathrm{Sp} = E_R^{-1/4}$ . In particular, the value $E_R=10^{-4}$ used for the low-amplitude analysis corresponds to $\mathrm{Sp} = 10$.

\begin{figure}
    \centering
    \includegraphics[width=0.95\textwidth]{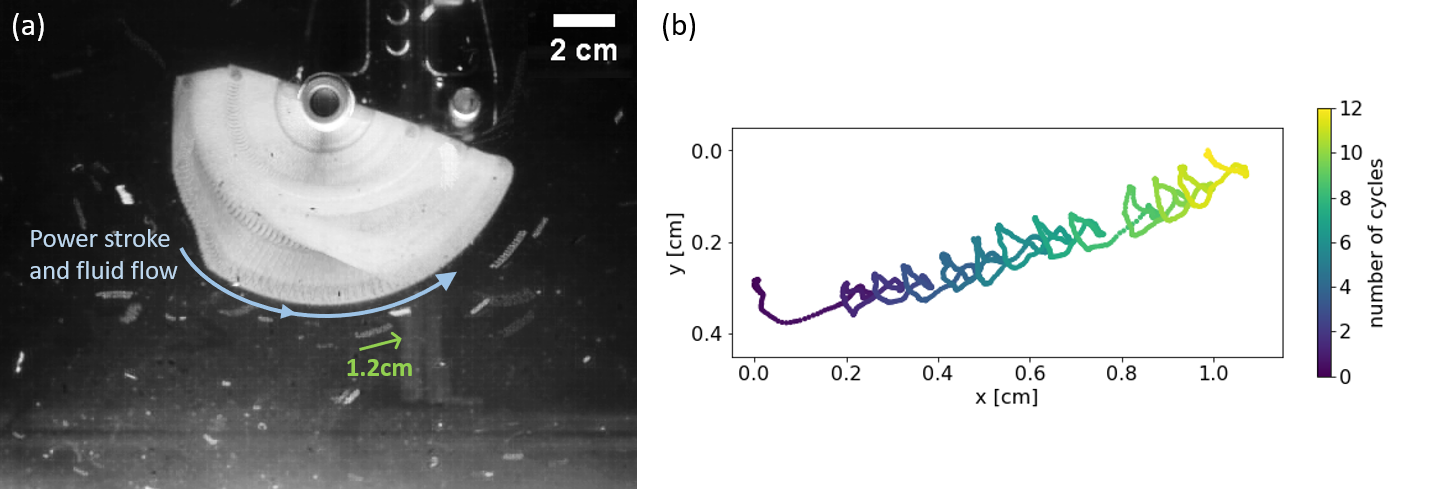}
    \caption{\bl (a) A maximum intensity projection of Supplementary Video 1, showing 12 beat cycles. The paths of the `tracer particles' can be seen as white streaks. The tracer particle annotated by a green arrow moves a total distance of 1.2~cm over the course of the video, so that the net flow speed is 0.1~cm per beat cycle. (b)~Detailed position of the annotated particle throughout the video, colour coded by time.}
    \label{fig:pumping}
\end{figure}

{\bl \section{Net fluid pumping by a robotic cilium}
Supplementary Video 1 shows 12 beat cycles of the robotic cilium beating in a tank of glycerin. The presence of small waste particles of cured resin in the tank allows the flow to be traced in certain locations. As we are interested in the average pumping rate, we perform a maximum intensity projection of the video, Fig \ref{fig:pumping}, so that the paths of the `tracer particles' show up as streaks. Visual inspection of the video confirms that the net movement of the tracers is in the same direction as the ciliary power stroke. The focal plane is sufficiently deep that many of the tracer particles are in the plane either above or below the beat plane of the robotic cilium. The strongest flow will be in the cilium beat plane, so we measure one of the longer tracer paths for this calculation, on the assumption that it is close to the beat plane. Over the 12 cycles of the video, this particle travels a net distance of 1.2~cm, giving an average net flow speed of 0.1~cm per beat cycle. }

\newpage

\bibliographystyle{abbrvnat}
\bibliography{bib}